\documentstyle[preprint,aps,tighten]{revtex}
\begin{document}
\draft
\title{Nonlocal calculation for nonstrange dibaryons and tribaryons}

\author{R. D. Mota $^{(1,2)}$, A. Valcarce $^{(3,4)}$,
F. Fern\'andez $^{(3)}$, D.R. Entem $^{(3,5)}$, and H. Garcilazo $^{(2)}$}

\address{$(1)$ Unidad Profesional Interdisciplinaria en Ingenier\'{\i}a, \\
Av. Instituto Polit\'ecnico Nacional No. 2580, Col. La Laguna Ticom\'an,\\
07740 M\'exico D.F., Mexico}

\address{$(2)$ Escuela Superior de F\' \i sica y Matem\'aticas \\
Instituto Polit\'ecnico Nacional, Edificio 9,
07738 M\'exico D.F., Mexico}

\address{$(3)$ Grupo de F\'\i sica Nuclear, Universidad de Salamanca,\\
E-37008 Salamanca, Spain}

\address{$(4)$ Departamento de F\'\i sica Te\'orica, \\
Universidad de Valencia, E-46100 Valencia, Spain}

\address{$(5)$ Department of Physics, \\
University of Idaho, Moscow, ID 83844, USA}
\maketitle

\begin{abstract}

We study the possible existence of nonstrange dibaryons and
tribaryons by solving the bound-state problem of the two- and three-body
systems composed of nucleons and deltas. The two-body systems are
$NN$, $N\Delta$, and $\Delta\Delta$, while the three-body systems
are $NNN$, $NN\Delta$, $N\Delta\Delta$, and $\Delta\Delta\Delta$.
We use as input the nonlocal
$NN$, $N\Delta$, and $\Delta\Delta$
potentials derived from the
chiral quark cluster model by means of the resonating
group method. We compare with previous results obtained
from the local version based on the Born-Oppenheimer approximation.
\end{abstract}

\vspace*{.5cm}
\noindent
Pacs: 14.20.Gk, 13.75.Cs, 14.20.Pt, 12.40.Yx

\newpage

\section{Introduction}

Systems without strangeness are those which involve only nucleons and
nonstrange mesons like the pion or the eta.
In a series of recent papers the suggestion has been made that it may be
possible to observe unstable  nonstrange two- and three-baryon states
corresponding
to the bound-state solutions of the various systems composed of nucleons
and deltas \cite{MOT1,MOT2,MOT3,MOT4}. These
are the systems $NN$, $N\Delta$, $\Delta\Delta$, $NNN$, $NN\Delta$,
$N\Delta\Delta$, and $\Delta\Delta\Delta$. The bound
states  involving one or more unstable particles will show up in nature
as dibaryon or tribaryon resonances. In the case of two-body
systems (dibaryons) they will decay mainly into two nucleons and either one
or two pions, while for the three-body case (tribaryons) they will decay
mainly into three nucleons and either one, two, or three pions.

In the previous calculations of our group \cite{MOT1,MOT2,MOT3,MOT4},
the Born-Oppenheimer approximation was used in order to obtain a local
potential for the baryon-baryon interactions ($NN$, $N\Delta$, $\Delta\Delta$).
In this work, we will overcome the Born-Oppenheimer approximation by working
directly with a nonlocal potential
derived within the resonating group method (RGM) formalism.
This method allows, once the Hilbert space for the
six-body problem has been fixed, to treat the
inter-cluster dynamics in an exact way.

In order to perform the $NNN$, $NN\Delta$, $N\Delta\Delta$, and
$\Delta\Delta\Delta$ calculations we will take
advantage of the experience gained in the three-nucleon
bound-state problem \cite{HARP,BERT}.
In that case one knows that the dominant configuration of
the system is that in which all particles are in S-wave states.
However, in order to get reasonable results for the binding energy, the
S-wave two-body amplitudes used as input in the Faddeev equations must
contain already the effect of the tensor force. Thus, for example, in the
case of the Reid soft-core potential if one considers only the S-wave
configurations but neglects the tensor force in the two-body subsystems
the triton is unbound. However, if one includes the effect of the
tensor force in the nucleon-nucleon $^3S_1$-$^3D_1$ channel, but uses
only the $^1S_0$ and $^3S_1$ components of the two-body
amplitudes in the three-body equations (2-channel calculation),
one gets a triton binding energy of 6.58 MeV. Notice
that including the remaining configurations (34-channel calculation),
leads to a triton binding energy of 7.35 MeV \cite{GLOC}. This means that
the S-wave truncated T-matrix approximation leads to a binding energy
which differs from the exact value by less than 1 MeV. Therefore,
by means of our approach we will not study exact binding
energies but which are the best candidates for bound states
and the ordering of the different $NNN$, $NN\Delta$,
$N\Delta \Delta$, and $ \Delta \Delta \Delta$
states.

$NN$, $N\Delta$, and $\Delta\Delta$ interactions have been
derived in the past in the framework of meson-exchange
models or phenomenological potentials  \cite{TER1,AR1}.
These models have been used over the
years to fit the $NN$ data very accurately. However,
in the $N\Delta$ and $\Delta\Delta$
sectors experimental data are so
scarce that it is not possible to obtain reliable
values of the parameters involved in the interaction.
The situation is different in the case of
quark cluster models \cite{FER1,FER2,DEN1}. In these models the
basic interaction is at the level of quarks involving only
a quark-quark-field (pion or gluon) vertex.
Therefore its parameters (coupling
constants, cutoff masses, etc.) are
independent of the
baryon to which the quarks are coupled, the difference among them being
generated by SU(2) scaling, as explained in Ref. \cite{BROW}.
Moreover, quark models provide a definite framework to
treat the short-range part of the interaction.
The Pauli principle between quarks determines the
short-range behavior of the different channels
without additional phenomenological assumptions.
In this way, even in the absence of experimental data,
one has a complete scheme which starting from the
$NN$ sector allows us to make predictions in the $N\Delta$
and $\Delta\Delta$ sectors. This fact is even more
important if one takes into account that the
short-range dynamics of the $N\Delta$ and $\Delta\Delta$
systems is to a large extent driven by
quark Pauli blocking effects, that
do not appear in the $NN$ sector.
Pauli blocking acts in a selective way in those channels
where the spin-isospin-color degrees of freedom are not enough to
accommodate all the quarks of the system \cite{FER3,FER4}.
Therefore, meson-exchange models cannot fully include the effect
of quark Pauli blocking through its purely phenomenological
short-range channel-independent part.

The lifetime of the bound states involving one or more deltas
should be similar to that of the $\Delta$ in the case of
very weakly bound systems and larger
if the system is very strongly bound. Therefore, these dibaryon
and tribaryon resonances will have widths similar or smaller
than the width of the $\Delta$ so that, in principle,
they are experimentally observable.
Also, we want to emphasize that the possible detection of
dibaryon and tribaryon resonances does not constitute an
exotic subject since, in principle, any nucleus with at least three nucleons
can serve as test system that may be
excited by forming a tribaryon \cite{HUBER}.

The paper is organized as follows.
In section II we present the basic quark-quark
interaction and we describe the method to obtain the
resonating group method baryon-baryon potentials.
Section III is dedicated to discuss the formalism
to solve the bound state problem for the cases of
systems of identical and non-identical particles, respectively.
In section IV we give our results and we
present the conclusions in section V.

\section{The two-body interactions}

The basic two-body interactions,
$V_{AB\to AB}$,
between baryons $A$ and $B$
that are going to be
used in this work are the nucleon-nucleon interaction $V_{NN\to NN}$,
the nucleon-delta interaction $V_{N\Delta\to N\Delta}$,
and the delta-delta
interaction $V_{\Delta\Delta\to \Delta\Delta}$.
These baryon-baryon interactions were
obtained from the chiral quark cluster model
developed elsewhere \cite{FER2}. In this model
baryons are described as clusters of
three interacting massive (constituent)
quarks, the mass coming from the breaking of chiral symmetry.
The ingredients of the quark-quark interaction are confinement,
one-gluon (OGE), one-pion (OPE) and one-sigma (OSE) exchange terms,
and whose parameters are fixed from
the $NN$ data. Explicitly,
the quark-quark ($qq$) interaction is,

\begin{equation}
V_{qq}(\vec r_{ij})= V_{\rm con} (\vec r_{ij}) +
V_{\rm OGE} (\vec r_{ij}) + V_{\rm OPE} (\vec r_{ij}) +
V_{\rm OSE} (\vec r_{ij}) \, ,
\label{term}
\end{equation}

\noindent
where ${\vec r}_{ij}$ is the $ij$ interquark distance and

\begin{equation}
V_{\rm con} ({\vec r}_{ij}) =
-a_c \, {\vec \lambda}_i \cdot {\vec
\lambda}_j \, r_{ij} \,  ,
\end{equation}

\begin{equation}
V_{\rm OGE} ({\vec r}_{ij}) =
{1 \over 4} \, \alpha_s \, {\vec
\lambda}_i \cdot {\vec \lambda}_j
\Biggl \lbrace {1 \over r_{ij}} -
{\pi \over m^2_q} \, \biggl [ 1 + {2 \over 3}
{\vec \sigma}_i \cdot {\vec
\sigma}_j \biggr ] \, \delta({\vec r}_{ij})
- {3 \over {4 m^2_q \, r^{3}_{ij}}}
\, S_{ij} \Biggr \rbrace \, ,
\label{OGE}
\end{equation}

\begin{eqnarray}
V_{\rm OPE} ({\vec r}_{ij}) & = & {1 \over 3}
\, \alpha_{ch} {\Lambda^2  \over \Lambda^2 -
m_\pi^2} \, m_\pi \, \Biggr\{ \left[ \,
Y (m_\pi \, r_{ij}) - { \Lambda^3
\over m_{\pi}^3} \, Y (\Lambda \,
r_{ij}) \right] {\vec \sigma}_i \cdot
{\vec \sigma}_j + \nonumber \\
 & & \left[ H( m_\pi \, r_{ij}) - {
\Lambda^3 \over m_\pi^3} \, H( \Lambda \,
r_{ij}) \right] S_{ij} \Biggr\} \,
{\vec \tau}_i \cdot {\vec \tau}_j \, ,
\label{OPE}
\end{eqnarray}

\begin{equation}
V_{\rm OSE} ({\vec r}_{ij}) = - \alpha_{ch} \,
{4 \, m_q^2 \over m_{\pi}^2}
{\Lambda^2 \over \Lambda^2 - m_{\sigma}^2}
\, m_{\sigma} \, \left[
Y (m_{\sigma} \, r_{ij})-
{\Lambda \over {m_{\sigma}}} \,
Y (\Lambda \, r_{ij}) \right] \, ,
\end{equation}

\noindent
where

\begin{equation}
Y(x) \, = \, {e^{-x} \over x} \,\,\,\,\, ; \,\,\,\,\,
H(x) \, = \, \Bigl( 1 + {3 \over x} + { 3 \over {x^2}} \Bigr) Y(x) \, .
\end{equation}

\noindent
Although taken to be linear for
consistency with the baryon and meson
spectra,
the detailed radial structure and strength
of the confining potential is
meaningless for the two-baryon interaction \cite{SHI}.
$a_c$ is the confinement strength, the ${\vec \lambda}$'s are the
SU(3) color matrices, the ${\vec \sigma}$'s
(${\vec \tau}$'s)
are the spin (isospin) Pauli matrices,
$S_{ij}$ is the usual tensor operator, $m_q$ ($m_\pi$, $m_\sigma$)
is the quark (pion, sigma) mass,
$\alpha_s$ is the $qq$-gluon coupling constant, $\alpha_{ch}$ is the
$qq$-meson coupling constant and $\Lambda$
is a cut-off parameter.

For the present study we make use of the nonlocal potentials
derived through a Lippmann-Schwinger formulation of the
RGM equations in momentum space \cite{DEN1}.
The formulation of the RGM for a system of two
baryons, $B_1$ and $B_2$, needs the wave function
of the two-baryon system constructed from
the one-baryon wave functions. The two-baryon
wave function can be written as:

\begin{equation}
\Psi_{B_1 B_2}={\cal A} [\chi(\vec{P}) \Psi_{B_1 B_2}^{ST} ]=
{\cal A} [\chi(\vec{P}) \phi_{B_1}(\vec{p}_{\xi_{B_1}}) \phi_{B_2}(\vec{p}_{\xi_{B_2}})
\chi_{B_1 B_2}^{ST} \xi_c [2^3]] ,
\label{rel}
\end{equation}

\noindent
where ${\cal A}$ is the antisymmetrizer of the six-quark system,
$\chi(\vec{P})$ is the relative wave-function of the two clusters,
$\phi_{B_1}(\vec{p}_{\xi_{B_1}})$ is the internal spatial
wave function of the baryon $B_1$, $\xi_{B_1}$ are the internal
coordinates of the three quarks of baryon $B_1$. $\chi_{B_1 B_2}^{ST}$
denotes the spin-isospin wave function of the two-baryon system coupled
to total spin ($S$) and isospin ($T$), and finally, $\xi_c [2^3]$
is the product of two color singlets.

The dynamics of the system is governed by the
Schr\"odinger equation
\begin{equation}
({\cal H} -E_T)| \Psi>=0 \Rightarrow <\delta \Psi | ({\cal H} -E_T)
| \Psi>=0,
\label{variations}
\end{equation}
where
\begin{equation}
{\cal H} = \sum_{i=1}^N {\vec{p_i}^2 \over 2 m_q} +\sum_{i<j} V_{ij}
-T_{c.m.}
\end{equation}
being $T_{c.m.}$ the center of mass kinetic energy, $V_{ij}$
the quark-quark interaction described above, and $m_q$
the constituent quark mass.

Assuming the functional form

\begin{equation}
\phi_B( \vec p ) = {\left( {b^2 \over \pi} \right)}^{3/4} e^{{-b^2 p^2}/2} \, ,
\end{equation}

\noindent
where $b$ is related to the size of the nucleon quark core,
Eq. (\ref{variations}) can be written in the following way,
\begin{equation}
\left ( {\vec{P}'^2 \over 2\mu} -E \right ) \chi(\vec{P}') +
\int K (\vec{P}', \vec{P}_i)
\chi(\vec{P}_i) d\vec{P}_i =0
\label{ALS}
\end{equation}

\noindent
where

\begin{equation}
K (\vec{P}', \vec{P}_i) =
{}^{RGM}V_D(\vec{P}', \vec{P}_i) + {}^{RGM}V_{EX}(\vec{P}', \vec{P}_i)
\label{ABLS}
\end{equation}
contains the direct and exchange RGM potentials, the later one coming from
quark antisymmetry.
$K(\vec{P}', \vec{P}_i)$ is the nonlocal potential.
>From Eq. (\ref{ALS}) a set of coupled Lippmann-Schwinger
equations can be obtained and solved using standard
techniques.
The parameters of the model are summarized in Table I.
They have been fixed in order to obtain the best fit of the
two-nucleon sector (deuteron binding energy and S-wave $NN$
scattering phase shifts) and the $\Delta -N$ mass difference.
In particular, the mass of the quark ($m_q$) is taken to be $1/3$
of the nucleon mass. The pion mass ($m_\pi$)
is its experimental value. The chiral coupling constant ($\alpha_{ch}$)
has been determined to reproduce the long-range OPE interaction and
is given by $\alpha_{ch} =\left( 3 \over 5 \right)^2 {g^2_{\pi NN}
\over {4 \pi}} {m^2_\pi \over {4 m^2_N}}$, where the $\pi NN$
coupling constant is taken to be ${g^2_{\pi NN} \over {4 \pi}} =$13.87.
The sigma mass is fixed by the chiral symmetry
relation $m^2_\sigma \approx
m^2_\pi + (2 m_q)^2$. The parameter $b$, which determines the size of
the nucleon quark content, was determined by comparing the adiabatic
$NN$ potential calculated from the wave function solution of the
bound state problem for the potential given by Eq. (1)
to the $NN$ potential calculated with a single gaussian of
parameter $b$. $\Lambda$, which controls the pion-gluon
proportion in the model and, as a consequence, the strength
of the tensor force, has been taken to reproduce the deuteron
binding energy in the presence of $\Delta \Delta$ channels.
As the OPE provides part of the $\Delta -N$ mass difference,
the value of the strong coupling constant ($\alpha_s$) is determined
to obtain the remaining $\Delta -N$ mass difference.
Finally, the value
of $a_c$, quoted for completeness because its contribution to the
baryon-baryon potential is negligible \cite{SHI},
is obtained from the stability condition for the
nucleon ${\partial M_N(b) \over \partial b} =0$ \cite{FER2}.

\section{Integral Equations}

We will describe in this section the formalism required in the cases of
the two-body systems $NN$, $N\Delta$, and $\Delta\Delta$ and
the three-body systems $NNN$, $NN\Delta$, $N\Delta\Delta$ and
$\Delta\Delta\Delta$.

\subsection{The two-body systems}

If we consider two baryons $A$ and $B$ in a relative
S-state interacting through a potential
that contains a tensor force, then there is a coupling to the $AB$
D-wave so that the Lippmann-Schwinger
equation of the system is of the form

\begin{eqnarray}
t_{i;j_ii_i}^{l_is_il_i^{\prime\prime}s_i^{\prime\prime}}
(p_i,p_i^{\prime\prime};E) & = &
V_{i;j_ii_i}^{l_is_il_i^{\prime\prime}s_i^{\prime\prime}}
(p_i,p_i^{\prime\prime}) + \sum_{l_i^{\prime}s_i^{\prime}}
\int_0^\infty {p_i^{\prime}}^2dp_i^{\prime}\,
V_{i;j_ii_i}^{l_is_il_i^{\prime}s_i^{\prime}}(p_i,p_i^{\prime}) \nonumber \\
& & \times {1 \over E - {p_i^{\prime}}^2/2\eta_i + i\epsilon}
t_{i;j_ii_i}^{l_i's_i'l_i^{\prime\prime}s_i^{\prime\prime}}
(p_i^{\prime},p_i^{\prime\prime};E),
\label{equ1}
\end{eqnarray}
where $j_i$ and $i_i$ are the  angular momentum and  isospin of the
system, while $l_is_i$, $l_i^{\prime}s_i^{\prime}$,
and $l_i^{\prime\prime}s_i^{\prime\prime}$
are the initial, intermediate, and
final orbital angular momentum and
spin of the system, respectively. $p_i$
and $\eta_i$ are, respectively, the
relative momentum and reduced mass of
the two-body system.
We give in Tables II, III and IV the corresponding $NN$, $N\Delta$
and $\Delta\Delta$
two-body channels in a relative S-wave that are coupled together for the
two possible values of $j$ and $i$
(since the $NN$ state is the one with the lowest mass, in the case of
this system we have considered also the possibility of transitions to
higher mass states like $N\Delta$ and $\Delta\Delta$).
In the cases of the $NN$ and $\Delta\Delta$ systems the Pauli principle
requires that $(-)^{l_i+s_i+i_i} = -1$.

As mentioned before, for the solution of
the three-body system we will use
only the component of the T-matrix obtained from the solution of
Eq. (\ref{equ1}) with $l_i=l_i^{\prime\prime}=0$,
so that for that purpose we define the
S-wave truncated amplitude

\begin{equation}
t_{i;s_ii_i}(p_i,p_i^{\prime\prime};E) \equiv
t_{i;s_ii_i}^{0s_i0s_i}(p_i,p_i^{\prime\prime};E).
\label{equ7}
\end{equation}

\subsection{The three-body systems}

If we restrict ourselves to the configurations where all three particles
are in S-wave states, the Faddeev equations for the bound-state problem
in the case of three particles with total spin $S$ and
total isospin $I$ are
\begin{eqnarray}
T_{i;SI}^{s_ii_i}(p_iq_i) & = & \sum_{j\ne i} \sum_{s_ji_j}
h_{ij;SI}^{s_ii_is_ji_j}{1 \over 2}
\int_0^\infty q_j^2 dq_j \int_{-1}^1 dcos\theta\,
t_{i;s_ii_i}(p_i,p^\prime_i;
E - q_i^2/2\nu_i) \nonumber \\
& & \times {1 \over E - p_j^2/2\eta_j -q_j^2/2\nu_j}\,
T_{j;SI}^{s_ji_j}(p_jq_j),
\label{for1}
\end{eqnarray}
where $p_i$ and $q_i$ are the usual Jacobi coordinates and $\eta_i$ and
$\nu_i$ the corresponding reduced masses

\begin{eqnarray}
\eta_i & = & {m_j m_k \over m_j + m_k},
\label{for2} \\
\nu_i & = & {m_i(m_j+m_k) \over m_i+m_j+m_k},
\label{for3}
\end{eqnarray}
with $ijk$ an even permutation of $123$. The momenta $p'_i$ and $p_j$
in Eq. (\ref{for1}) are given by

\begin{eqnarray}
{p^\prime_i}^2 & = & q_j^2 + {\eta_i^2 \over m_k^2}q_i^2
+ 2{\eta_i \over m_k}q_i q_j cos\theta,
\label{for4} \\
p_j^2 & = & q_i^2 + {\eta_j^2 \over m_k^2}q_j^2
+ 2{\eta_j \over m_k}q_i q_j cos\theta.
\label{for5}
\end{eqnarray}
$h_{ij;SI}^{s_ii_is_ji_j}$ are the spin-isospin coefficients

\begin{eqnarray}
h_{ij;SI}^{s_ii_is_ji_j} & = & (-)^{s_j+\sigma_j-S}\sqrt{(2s_i+1)(2s_j+1)}\,
W(\sigma_j \sigma_k S \sigma_i;s_i s_j) \nonumber \\
& & \times (-)^{i_j+\tau_j-I}\sqrt{(2i_i +1)(2i_j +1)}\,
W(\tau_j \tau_k I \tau_i;i_i i_j),
\label{for6}
\end{eqnarray}
where $W$ is the Racah coefficient
and $\sigma_i$, $s_i$, and $S$ ($\tau_i$,
$i_i$, and $I$) are the spin (isospin) of particle $i$, of
the pair $jk$, and of the three-body system, respectively.

Since the variable $p_i$, in Eqs. (\ref{equ1}) and
(\ref{for1}), runs from $0$ to $\infty$,
it is convenient to make the transformation

\begin{equation}
x_i = {p_i - d \over p_i + d},
\label{for9}
\end{equation}
where the new variable $x_i$ runs from $-1$ to $1$, and $d$ is a scale
parameter. With this transformation Eq. (\ref{for1}) takes the form

\begin{eqnarray}
T_{i;SI}^{s_ii_i}(x_iq_i) & = & \sum_{j\ne i} \sum_{s_ji_j}
h_{ij;SI}^{s_ii_is_ji_j}{1 \over 2}
\int_0^\infty q_j^2 dq_j \int_{-1}^1 dcos\theta\,
t_{i;s_ii_i}(x_i,x^\prime_i;
E - q_i^2/2\nu_i) \nonumber \\
& & \times {1 \over E - p_j^2/2\eta_j -q_j^2/2\nu_j}\,
T_{j;SI}^{s_ji_j}(x_jq_j).
\label{for10}
\end{eqnarray}
Since in the amplitude $t_{i;s_ii_i}(x_i,x^\prime_i;e)$
the variables $x_i$ and ${x'}_i$
run from $-1$ to $1$, one can expand this amplitude in terms of Legendre
polynomials as

\begin{equation}
t_{i;s_ii_i}(x_i,x^\prime_i;e)=\sum_{nm}P_n(x_i)
\tau_{i;s_ii_i}^{nm}(e)P_m(x^\prime_i),
\label{for11}
\end{equation}
where the expansion coefficients are given by

\begin{equation}
\tau_{i;s_ii_i}^{nm}(e)={2n+1 \over 2}\,{2m+1 \over 2}
\int_{-1}^1 dx_i \int_{-1}^1
dx^\prime_i\, P_n(x_i)t_{i;s_ii_i}(x_i,x^\prime_i;e)
P_m(x^\prime_i).
\label{for12}
\end{equation}
Applying expansion (\ref{for11}) in Eq. (\ref{for10}) one gets

\begin{equation}
T_{i;SI}^{s_ii_i}(x_iq_i) = \sum_n T_{i;SI}^{ns_ii_i}(q_i)P_n(x_i),
\label{for13}
\end{equation}
where $T_{i;SI}^{ns_ii_i}(q_i)$ satisfies the one-dimensional
integral equation

\begin{equation}
T_{i;SI}^{ns_ii_i}(q_i)=\sum_{j\ne i}\sum_{ms_ji_j} \int_0^\infty dq_j\,
A_{ij;SI}^{ns_ii_ims_ji_j}(q_i,q_j;E)T_{j;SI}^{ms_ji_j}(q_j),
\label{for14}
\end{equation}
with

\begin{eqnarray}
A_{ij;SI}^{ns_ii_ims_ji_j}(q_i,q_j;E) & = & h_{ij;SI}^{s_ii_is_ji_j}  \sum_l
\tau_{is_ii_i}^{nl}(E-q_i^2/2\nu_i)
{q_j^2 \over 2} \nonumber \\
& & \times \int_{-1}^1 dcos\theta\,{P_l(x_i)P_m(x_j) \over E-p_j^2/2\eta_j
-q_j^2/2\nu_j}.
\label{for15}
\end{eqnarray}

The three amplitudes $T_{1;SI}^{ls_1i_1}(q_1)$, $T_{2;SI}^{ms_2i_2}(q_2)$,
and
$T_{3;SI}^{ns_3i_3}(q_3)$ in Eq. (\ref{for14}) are coupled together.
The number of
coupled equations can be
reduced, however, since some of the particles are identical.
In the case of three identical particles ($NNN$ and
$\Delta\Delta\Delta$ systems) we have that all three amplitudes are
equal and therefore Eq. (\ref{for14}) becomes in this case,

\begin{equation}
T_{SI}^{ns_ii_i}(q_i)=2\sum_{ms_ji_j} \int_0^\infty dq_j\,
A_{ij;SI}^{ns_ii_ims_ji_j}(q_i,q_j;E)T_{SI}^{ms_ji_j}(q_j).
\label{for16}
\end{equation}
We give in Table V the three $NNN$ states characterized by total spin
and isospin $(S,I)$ that are possible as well as the two-body
$NN$ channels that contribute to each state. In Table VI we give
the 25 $\Delta\Delta\Delta$ states
characterized by total spin and isospin
$(S,I)$ that are possible as well as the two-body $\Delta\Delta$
channels that contribute to each state.

In the case where two particles are identical and one different
($NN\Delta$ and $N\Delta\Delta$ systems) two of the amplitudes are equal.
The reduction procedure for the case where one has two
identical fermions has been described before \cite{AFN1,GAMI}
and will not be repeated
here. With the assumption that particle 1 is the different one and
particles 2 and 3 are the two identical, only the amplitudes
$T_{1;SI}^{ns_1i_1}(q_1)$ and $T_{2;SI}^{ms_2i_2}(q_2)$ are
independent from each other and
they satisfy the coupled integral equations

\begin{eqnarray}
T_{1;SI}^{ls_1i_1}(q_1) & = &2 \sum_{ns_2i_2} \int_0^\infty dq_3\,
A_{13;SI}^{ls_1i_1ns_2i_2}(q_1,q_3;E)
T_{2;SI}^{ns_2i_2}(q_3),
\label{for16p} \\
T_{2;SI}^{ms_2i_2}(q_2) & = & \sum_{ns_3i_3} (-)^{Iden}\int_0^\infty
dq_3\,A_{23;SI}^{ms_2i_2ns_3i_3}(q_2,q_3;E)
T_{2;SI}^{ns_3i_3}(q_3)  \nonumber \\
 & &   + \sum_{ls_1i_1} \int_0^\infty dq_1\,
 A_{31;SI}^{ms_2i_2ls_1i_1}(q_2,q_1;E)
 T_{1;SI}^{ls_1i_1}(q_1),
 \label{for17}
\end{eqnarray}
with the identical-particles phase

\begin{equation}
Iden = 1+\sigma_1 +\sigma_3-s_2 + \tau_1+\tau_3-i_2.
\label{for18}
\end{equation}
Substitution of Eq. (\ref{for16p}) into Eq. (\ref{for17}) yields an
equation with only the amplitude $T_2$

\begin{equation}
T_{2;SI}^{ms_2i_2}(q_2)  =  \sum_{ns_3i_3} \int_0^\infty
dq_3\,K_{23;SI}^{ms_2i_2ns_3i_3}(q_2,q_3;E)
T_{2;SI}^{ns_3i_3}(q_3),
\label{for19}
\end{equation}
where

\begin{eqnarray}
K_{23;SI}^{ms_2i_2ns_3i_3}(q_2,q_3;E) & = & (-)^{Iden}
A_{23;SI}^{ms_2i_2ns_3i_3}(q_2,q_3;E)+2\sum_{ls_1i_1}
\int_0^\infty dq_1\, \nonumber \\
& & \times A_{31;SI}^{ms_2i_2ls_1i_1}(q_2,q_1;E)
A_{13;SI}^{ls_1i_1ns_3i_3}(q_1,q_3;E).
\label{for20}
\end{eqnarray}

We give in Table VII the 9 $NN\Delta$
states characterized by total spin and isospin $(S,I)$ that are possible
as well as the two-body $N\Delta$ and $NN$ channels that contribute
to each state. In Table VIII we give the 16
$N\Delta\Delta$ states characterized by total spin and isospin $(S,I)$
that are possible as well as the two-body $N\Delta$ and
$\Delta\Delta$ channels that contribute to each state.

\subsection{Numerical solutions}

In order to find the bound-state solutions of Eqs.
(\ref{equ1}), (\ref{for16}) and (\ref{for19}) we drop the inhomogeneous
term in Eq. (\ref{equ1}) (of course, in the solution of the three-body
problem we use as input the solutions of the inhomogeneous Eq.
(\ref{equ1})) and
replace the
integral by a
sum applying a numerical integration quadrature \cite{ABRA}.
In this way,
Eqs. (\ref{equ1}), (\ref{for16}) and (\ref{for19})
become a set of homogeneous linear equations. This set of
linear equations has solutions only if the determinant of the matrix
of the coefficients (the Fredholm determinant) vanishes for certain
energies. Thus, the procedure to find the bound states of the system
consists simply in searching for the zeroes of the Fredholm
determinant as a function of energy.
We checked our program by comparing with known results for the three-nucleon
bound-state problem with the Reid soft-core potential \cite{HARP}.
We found very
stable results taking for the scale parameter $d$ = 3 fm$^{-1}$, a
number of Legendre polynomials $L=10$, and a number of Gauss-Legendre
points $N=12$.

\section{Results}

We will now present the results of our nonlocal calculations for the 7
systems  corresponding to the two- and
three-body bound-state problem of nucleons and deltas, and compare
them to previous calculations which have been done by our group
based on the local potentials obtained from the Born-Oppenheimer
approximation.

The two body interaction in the $N\Delta$ states $(j,i)=$ (1,1)
and (2,2), and those of the $\Delta \Delta$ states $(j,i)=$ (2,3)
and (3,2) present quark Pauli blocking. As a consequence,
a strong repulsive core appears in the baryon-baryon potential.
The reason for that is based on the fast decrease of the norm
of the six-quark wave function when $R \to 0$ \cite{FER3}. A similar
analysis performed in terms of the SU(4) symmetry shows the
presence of a forbidden state. From the physical
point of view, it is connected with the lack of enough degrees of
freedom to accommodate all the quarks. It is important to note
that the origin of this repulsion is not the same as in the $NN$
channels, because they do not show a forbidden state but a
mixing of $[6]$ with the $[4,2]$ six-quark orbital symmetry.
Technically, the reason for such a strong repulsive core
is the presence of nodes in the inner
region of the relative wave function of Eq. (\ref{rel}).
This behavior originates essentially from the condition that the
relative wave function should be orthogonal to the forbidden
state due to the Pauli principle \cite{SAITO}. The forbidden state
should then be eliminated from the relative wave function for each
partial wave. This procedure is tedious both
from the conceptual and numerical point of view \cite{SAITO,OTSUKI}.
It has been demonstrated \cite{TOKI} that for the Pauli blocked channels
the local $N\Delta$ and $\Delta\Delta$ potentials reproduce
the qualitative behavior of the RGM kernels after the
subtraction of the forbidden states. This is why
we used in our calculations the local version of the
quark Pauli blocked channels mentioned above.

In the case of the three-body systems we calculated the binding-energy
spectrum (that is, the energy of the states measured with respect  to the
three-body threshold) as well as the separation-energy spectrum
(that is, the energy of the states measured with respect  to the threshold
of one free particle and a bound state of the other two). The deepest
bound three-body state is not the one with the largest binding energy
but the one with the largest separation energy, since that state is the
one that requires more energy in order to become unbound
(that is, to move  it from the bound state to the nearest threshold).

\subsection{The $NN$ system}

We found that of the two states of Table II only the one with
$(j,i)=(1,0)$, that is the deuteron, is bound.
The nonlocal model gives a deuteron binding energy of
2.14 MeV, while the local version gave an energy of 3.13 MeV.
These results are shown in Table IX.

The  exact chiral quark cluster
model $NN$ potential \cite{DEN1} gives
a deuteron binding energy of 2.225 MeV. This value was obtained
by taking into account the $\Delta\Delta$ partial wave
$(l_{\Delta\Delta},s_{\Delta\Delta})=(4,3)$
coupled together in addition to those
given in Table II. Since in our calculation we consider
only S- and D-waves, we omit
the $\Delta\Delta$ $(l_{\Delta\Delta},s_{\Delta\Delta})=(4,3)$
partial-wave contribution, and
we obtain instead a deuteron binding energy of 2.14 MeV, which differs
less than 0.1 MeV from the exact calculation.

\subsection{The $N\Delta$ system}

We give in Table X the results for the binding energies of the
$N\Delta$ system. Out of the four possible
$N\Delta$ states of Table III
only one, the $(j,i)=(2,1)$, has a bound state which lies exactly
at the $N\Delta$ threshold for the local model. However, if we use
the nonlocal model we find instead
a bound state of 0.141 MeV.
The states $(j,i)=(1,1)$ and $(2,2)$ are unbound
because they present
quark Pauli blocking \cite{FER3} and therefore
they have a strong repulsive barrier at short
distances in the S-wave central interaction.
These two states play an important role in the
three-body spectrum.
The state $(j,i)=(2,1)$ can also exist
in the $NN$ system and there it corresponds to the $^1D_2$ partial wave
which has a resonance at an invariant mass of 2.17 GeV
\cite{HOSH,YOKO,ARND}.
This means that the $N\Delta$ bound state may decay into two nucleons
and appear in the $NN$ system as a resonance.
The $N\Delta$ bound state has for both local and nonlocal models
energies very close to the
$N\Delta$ threshold, so that
the invariant mass of the system is also very close
to 2.17 GeV. Thus, one or another
of our models
predict the $NN$ $^1D_2$ resonance as being a $N\Delta$ bound state.

\subsection{The $\Delta\Delta$ system}

We give in Table XI our results for the $\Delta\Delta$ system. Out of
the eight possible $\Delta\Delta$ states given in Table IV
with nonlocal interactions five have a bound state,
whereas the
local interactions bind six of them
(in both local and nonlocal models there are no excited states in any
of the channels). It is interesting
to note that the predicted bound states:
$(j,i)=(1,0)$, $(0,1)$, $(2,1)$ and $(3,0)$,
also appear in the case
of the $NN$ system. In the nonlocal model, we find that the deepest
bound state is $(j,i)=(1,0)$, the second $(j,i)=(0,1)$, the
third $(j,i)=(3,0)$ and the fourth $(j,i)=(2,1)$.
This clearly shows that there is a qualitative
similarity between the $\Delta\Delta$ and $NN$ systems (both are
systems of identical particles).
Three of these
states appear also in the case of the $NN$ system. The $(j,i)=(1,0)$
state is of course the deuteron, the $(j,i)=(0,1)$ is the
$^1S_0$ virtual state and the $(j,i)=(2,1)$ state is the
$^1D_2$ resonance that lies at $\approx$ 2.17 GeV \cite{HOSH}
(note that the $^3F_3$ $NN$ resonance has no counterpart in
Table XI because we calculated only even-parity states and
$^3F_3$ has odd parity). Thus, the $(j,i)=(3,0)$ state, which is
also allowed in the case of the $NN$ system, would correspond to a
new nucleon-nucleon resonance that is predicted by our model.
The $(j,i)=(3,0)$ channel corresponds in the case of the
$NN$ system to the $^3D_3$ partial wave.
Some indication of the $(3,0)$ resonance can already
be seen in the most recent analysis of the $NN$ data by
Arndt {\it et al.} \rm \cite{ARND}.

The channels $(j,i)=(2,3)$ and $(3,2)$ are unbound because they have a
strong repulsive barrier at short distances in the S-wave central
interaction. This strong repulsion originates from the quark Pauli
blocking produced by the saturation of states that occurs when the
total spin and isospin are near their maximum values \cite{FER4}. As we will
see later in the discussion of the $\Delta\Delta\Delta$ results, these
repulsive cores in the $(3,2)$ and $(2,3)$ channels largely determine
the three-body spectrum.

>From Table XI we note that the two-body $\Delta\Delta$ bound states
which have low quantum
numbers are deeper for the nonlocal model than with the local one.
This peculiar feature results to be conversely
for the case of high quantum numbers.

\subsection{The $NNN$ system}

As another test of the reliability of our model in the case of the
three-baryon system we solved the $NNN$ bound-state problem. We found
that of the states of Table V only the state with
$(S,I)=({1\over 2},{1\over 2})$, that is the triton, has a bound
state. By using the local potentials we obtain a binding energy of 5.76
MeV for the triton. On other hand, if we use the nonlocal potentials
as input we find a triton binding energy of 6.52 MeV.
For comparison, we notice that
the triton binding energy for the Reid-soft-core potential in the
truncated T-matrix approximation is 6.58 MeV.
Since the experimental value is $B_{EXP}=8.49$ MeV the difference with
our theoretical result, of about 3 MeV,
is a measure of
the uncertainty of our calculation in the case of the three-baryon system.
We show in Table XII the results of our calculations for the $NNN$
system. There, $B_3$ is the binding energy of the system and $B_3-B_2$ is the
separation energy, being $B_2$ the binding energy of the deepest
bound two-body channel that contributes to the three-body state (see Table
IX).

\subsection{The $NN\Delta$ system}

We show in Table XIII the results of our calculations for the $NN\Delta$
system.

One may have hoped to find several bound
states in this system, due to
the fact that the $N\Delta$ two-body subsystem has a bound state in the
channel $(j,i)=(2,1)$ and the $NN$
two-body subsystem has a bound state
in the channel $(j,i)=(1,0)$ and
an almost-bound state in the channel
$(j,i)=(0,1)$. This is not the
case however, and as a matter of fact, with the nonlocal potentials as
input only two of the 9 possible three-body states
given in Table VII are bound.
Because of the attractive
contribution of the $N\Delta$ $(j,i)=(2,1)$ bound state with the
nonlocal model, the three-body state  $({3\over 2},{1\over 2})$
results to be very weakly bound, at an energy of 0.143 MeV, and a
separation energy scarcely different from zero.
That means that the $(S,I)=({3\over 2},{1\over 2})$ state is very near
the $NN\Delta$ threshold and therefore it represents the tribaryon
resonance with the lowest possible mass since it can decay into
three nucleons and one pion.
Also, for this case
the three-body state $({3\over 2},{3\over 2})$ is bound. As it can be
seen  from Table VII, this state has the contribution of all the
two-body $N\Delta$ and $NN$ channels. In spite of the fact that
the $N\Delta$ two-body channels $(j,i)=(1,1)$ and
$(2,2)$ present Pauli blocking \cite{FER3},
and therefore they have a strong repulsive
barrier at short distances in the S-wave central interaction,
the attractive contribution of
the $N\Delta$ $(j,i)=(2,1)$ and $NN$ $(j,i)=(1,0)$ channels
results to be enough to weakly bound this state.
We note that neither one of the three-body states
$(S,I)=({3\over 2},{1\over 2})$ and $({3\over 2},{3\over 2})$
is bound with local interactions.

\subsection{The $N\Delta\Delta$ system}

The results for the $N\Delta\Delta$ system are shown in Table XIV.
Similarly to the case just discussed,
in our calculations with nonlocal interactions
we found that three of the 16 possible
$N\Delta\Delta$ states given in Table VIII are bound.
They are the $(S,I)=({1 \over 2},{5\over 2})$,$({5\over 2},{1\over 2})$,
and $({5 \over 2},{5 \over 2})$ states and their corresponding bound
state energies are 0.630 MeV, 8.158 MeV, and 0.181 MeV, respectively.
In the case of the states $(S,I)=({1\over 2},{5\over 2})$
and $(S,I)=({5\over 2},{1\over 2})$ the repulsive barrier due
the quark Pauli blocking in
the $N\Delta$ states
$(j,i)=(1,1)$ and $(2,2)$ is less strong than the
attraction due to the state $(j,i)=(2,1)$, so that they result
to be bound states in the nonlocal model. The state $(S,I)=
({5\over 2},{5\over 2})$ is the weakest bound state of this system,
since in addition to the contribution of the $N\Delta$ quark Pauli
blocking channels, there exists that of the
$\Delta\Delta$ quark Pauli blocking channels $(j,i)=(2,3)$
and $(3,2)$. This confirms what we have mentioned
before that it is the structure of the interaction of the
two-body system the one which largely determines the three-body
spectrum.
Thus, the nonlocal interactions predict the bound states
$(S,I)=({1 \over 2},{5\over 2})$, $({5\over 2},{1\over 2})$
and
$({5\over 2},{5\over 2})$, which in principle may
be observable as tribaryon resonances which decay into three
nucleons and two pions
with masses close to the $N\Delta\Delta$ threshold.

\subsection{The $\Delta\Delta\Delta$ system}

We show in Table XV the results for the $\Delta\Delta\Delta$ system.
The system has 4 bound states while by using the local interactions
the system had 7 bound states.
>From Table XV we observe that the three states which are missing
in the nonlocal version
are barely bound in the local version, i.e., they have very small
separation energies. Since the nonlocal interaction tends to lower
the attraction in all the $\Delta\Delta\Delta$ channels it is not
surprising that those which were barely bound have now disappeared.
The more strongly bound three-body state
(that is, the one with the largest separation energy) is the
$(S,I)=({1\over 2},{1\over 2})$ state which has precisely
the quantum numbers
of the triton. This shows again, like in the $\Delta\Delta$
and $NN$ systems, the similarity
between the $\Delta\Delta\Delta$ and $NNN$ systems.

The reason why the $(S,I)=({1\over 2},{1\over 2})$ state is the more
strongly bound is very simple. As shown in Table VI, this is the only
state where none of the two-body channels with a strong repulsive core
$(j,i)=(2,3)$ or $(3,2)$ contribute. In all the other three-body states the
strong repulsion of the $(j,i)=(2,3)$ and $(3,2)$ channels either
completely destroys the bound state or allows just a barely bound one.
The state $(S,I)=({7\over 2},{3\over 2})$ comes next with respect to
separation energy. This state $(S,I)=({7\over 2},{3\over 2})$ has
a somewhat anomalous
behavior since it has a relatively large separation energy.
This behavior is sort of accidental and it
can be understood as follows. As seen in Table VI, there are 4 two-body
channels contributing to the $(S,I)=({7\over 2},{3\over 2})$ state, the
two attractive ones $(j,i)=(2,1)$ and $(3,0)$ and the two repulsive ones
$(j,i)=(2,3)$ and $(3,2)$. However, as one can see in Table XI
the attractive channels $(2,1)$ and $(3,0)$
have bound states at $E=-7.4$ MeV and $E=-7.8$ MeV, respectively,
for the nonlocal version, and $E=-30.5$ MeV and $E=-29.9$ MeV,
respectively, for the local version, so that
the poles in the scattering amplitudes of these two channels are very
close together and therefore there is a reinforcement between
them, which gives rise to the anomalously large separation energy
in both versions.

\section{Conclusions}

By using both the local and nonlocal models
we have studied the bound-state solutions of the two- and three-body systems
composed of nucleons and deltas.
First of all we would like to emphasize the goodness of the
Born-Oppenheimer approximation, producing results very similar
to the usually more involved RGM results. We
conclude that the more realistic nonlocal interactions produce in the
two-body systems $NN$, $N\Delta$, and $\Delta\Delta$ one, one,
and five bound states respectively.
The bound states of the unstable systems $N\Delta$ and $\Delta\Delta$
correspond to
dibaryon resonances that decay mainly into two nucleons and one pion and
two nucleons and two pions, respectively. The $N\Delta$ bound state
with $(j,i)=(2,1)$ and $M \approx 2.17$ GeV is the dibaryon resonance
with the lowest possible mass and the one which seems to be
well confirmed by experiment. The five $\Delta\Delta$ bound states
of the nonlocal potentials correspond to dibaryon resonances
with masses between 2.4 and 2.5 GeV. The $(j,i)=(3,0)$ $\Delta\Delta$
state would correspond to a new nucleon-nucleon resonance predicted
by our model. A possible signal of this resonance appears in a recent
analysis of $NN$ data up to 3 GeV by Arndt {\it et al.} \cite{ARND}.
With respect to the three-body systems we found that the $NNN$
has one bound state, the
$\Delta\Delta\Delta$ has four bound states, the $NN\Delta$ has
two bound states, and the $N\Delta\Delta$ has three bound
states.
The predicted $NN\Delta$ states with
$(S,I)=({3\over 2},{1\over 2})$ and $(S,I)=({3\over 2},{3\over 2})$
which correspond to $M \approx 3.4$ GeV are the
tribaryon resonances with the lowest mass and therefore the ones that
would be more easy to detect experimentally.

\acknowledgements
A.V. thanks the
hospitality of the Instituto Polit\'ecnico Nacional
of Mexico during several visits
when part of this work has been done. He also
thanks the Ministerio de Educaci\'on, Cultura y Deporte of Spain
for financial support through the Salvador de Madariaga program.
This work has been partially funded by
COFAA-IPN (Mexico),
by Direcci\'{o}n General de
Investigaci\'{o}n Cient\'{\i}fica y T\'{e}cnica (DGICYT) under the Contract
number PB97-1401, and by Junta de Castilla y Le\'on under the Contract number
SA-109/01, by the University of Salamanca and by the Ram\'on Areces
Foundation (Spain).

\begin{table}
\caption{ Quark model parameters.}

\begin{tabular}{cccc}
 & $m_q ({\rm MeV})$                   &  313     & \\
 & $b ({\rm fm})$                      &  0.518   & \\
\tableline
 & $\alpha_s$                    &  0.498  & \\
 & $a_c ({\rm MeV} \cdot {\rm fm}^{-1})$     &  67.0  & \\
 & $\alpha_{ch}$                 &  0.027 & \\
 & $m_\sigma ({\rm fm}^{-1})$          &  3.513 & \\
 & $m_\pi ({\rm fm}^{-1})$             &  0.70    & \\
 & $\Lambda ({\rm fm}^{-1})$           &  4.3     & \\
\end{tabular}
\end{table}

\begin{table}
\caption{$NN$ channels $(l_{NN},s_{NN})$,
$N\Delta$ channels $(l_{N\Delta},s_{N\Delta})$, and
$\Delta\Delta$ channels $(l_{\Delta\Delta},s_{\Delta\Delta})$
that are coupled together in the
$^3S_1$-$^3D_1$ and $^1S_0$ $NN$ states.}

\begin{tabular}{cccccc}
$NN$ state & $j$ & $i$ & $(l_{NN},s_{NN})$ & $(l_{N\Delta},s_{N\Delta})$
& $(l_{\Delta\Delta},s_{\Delta\Delta})$ \\
\tableline
$^3S_1$-$^3D_1$   & 1   &  0   &(0,1),(2,1)&   $-$  & (0,1),(2,1),(2,3) \\
$^1S_0$           & 0   &  1   &(0,0)      &(2,2) & $-$
\end{tabular}
\end{table}

\begin{table}
\caption{Coupled channels $(l,s)$ that contribute to a given
$N\Delta$ state with
total angular momentum $j$ and isospin $i$.}

\begin{tabular}{ccccc}
& $j$ & $i$   & $(l,s)$ & \\
\tableline
& 1   &  1    & (0,1),(2,1),(2,2) & \\
& 1   &  2    & (0,1),(2,1),(2,2) & \\
& 2   &  1    & (0,2),(2,1),(2,2) & \\
& 2   &  2    & (0,2),(2,1),(2,2) & \\
\end{tabular}
\end{table}

\begin{table}
\caption{Coupled channels $(l,s)$ that contribute to a given
$\Delta\Delta$ state with
total angular momentum $j$ and isospin $i$.}

\begin{tabular}{ccccc}
& $j$ & $i$   & $(l,s)$ & \\
\tableline
& 0   &  1    & (0,0),(2,2) & \\
& 0   &  3    & (0,0),(2,2) & \\
& 1   &  0    & (0,1),(2,1),(2,3) & \\
& 1   &  2    & (0,1),(2,1),(2,3) & \\
& 2   &  1    & (0,2),(2,0),(2,2) & \\
& 2   &  3    & (0,2),(2,0),(2,2) & \\
& 3   &  0    & (0,3),(2,1),(2,3) & \\
& 3   &  2    & (0,3),(2,1),(2,3) & \\
\end{tabular}
\end{table}

\begin{table}
\caption{Two-body $NN$ channels $(j,i)$ that contribute to a given
$NNN$ state with total spin $S$ and isospin $I$.}

\begin{tabular}{ccccc}
 & $S$ & $I$   & $(j,i)$ &  \\
\tableline
 & 1/2    &  1/2    & (1,0),(0,1) &   \\
 & 1/2    &  3/2    & (0,1) & \\
 & 3/2    &  1/2    & (1,0) & \\
\end{tabular}
\end{table}

\begin{table}
\caption{Two-body $\Delta\Delta$ channels $(j,i)$
that contribute to a given
$\Delta\Delta\Delta$ state with
total spin $S$ and isospin $I$.}

\begin{tabular}{ccccc}
& $S$ & $I$   & $(j,i)$ & \\
\tableline
& 1/2    &  1/2    & (1,2),(2,1) & \\
& 1/2    &  3/2    & (1,0),(1,2),(2,1),(2,3) & \\
& 1/2    &  5/2    & (1,2),(2,1),(2,3) & \\
& 1/2    &  7/2    & (1,2),(2,3) & \\
& 1/2    &  9/2    & (2,3) & \\
& 3/2    &  1/2    & (0,1),(1,2),(2,1),(3,2) & \\
& 3/2    &  3/2    & (0,1),(0,3),(1,0),(1,2), & \\
&        &         & (2,1),(2,3),(3,0),(3,2) & \\
& 3/2    &  5/2    & (0,1),(0,3),(1,2),(2,1), & \\
&        &         & (2,3),(3,2) & \\
& 3/2    &  7/2    & (0,3),(1,2),(2,3),(3,2) & \\
& 3/2    &  9/2    & (0,3),(2,3) & \\
& 5/2    &  1/2    & (1,2),(2,1),(3,2) & \\
& 5/2    &  3/2    & (1,0),(1,2),(2,1),(2,3), & \\
&        &         & (3,0),(3,2) & \\
& 5/2    &  5/2    & (1,2),(2,1),(2,3),(3,2) & \\
& 5/2    &  7/2    & (1,2),(2,3),(3,2) & \\
& 5/2    &  9/2    & (2,3) & \\
& 7/2    &  1/2    & (2,1),(3,2) & \\
& 7/2    &  3/2    & (2,1),(2,3),(3,0),(3,2) & \\
& 7/2    &  5/2    & (2,1),(2,3),(3,2) & \\
& 7/2    &  7/2    & (2,3),(3,2) & \\
& 7/2    &  9/2    & (2,3) & \\
& 9/2    &  1/2    & (3,2) & \\
& 9/2    &  3/2    & (3,0),(3,2) & \\
& 9/2    &  5/2    & (3,2) & \\
& 9/2    &  7/2    & (3,2) & \\
& 9/2    &  9/2    &  & \\
\end{tabular}
\end{table}

\begin{table}
\caption{Two-body $N\Delta$ channels $(j_{N\Delta},i_{N\Delta})$ and two-body
$NN$ channels $(j_{NN},i_{NN})$ that contribute to a given
$NN\Delta$ state with
total spin $S$ and isospin $I$.}
\label{NND}

\begin{tabular}{cccc}
 $S$ & $I$   & $(j_{N\Delta},i_{N\Delta})$ &  $(j_{NN},i_{NN})$ \\
\tableline
 1/2    &  1/2    & (1,1) &  \\
 1/2    &  3/2    & (1,1),(1,2) & (1,0) \\
 1/2    &  5/2    & (1,2) &  \\
 3/2    &  1/2    & (1,1),(2,1) & (0,1) \\
 3/2    &  3/2    & (1,1),(1,2),(2,1),(2,2) & (1,0),(0,1) \\
 3/2    &  5/2    & (1,2),(2,2) & (0,1) \\
 5/2    &  1/2    & (2,1) &  \\
 5/2    &  3/2    & (2,1),(2,2) & (1,0) \\
 5/2    &  5/2    & (2,2) &  \\
\end{tabular}
\end{table}

\begin{table}
\caption{Two-body $N\Delta$ channels $(j_{N\Delta},i_{N\Delta})$ and two-body
$\Delta\Delta$ channels $(j_{\Delta\Delta},i_{\Delta\Delta})$
that contribute to a given
$N\Delta\Delta$ state with
total spin $S$ and isospin $I$.}

\begin{tabular}{cccc}
 $S$ & $I$ & $(j_{N\Delta},i_{N\Delta})$ & $(j_{\Delta\Delta},i_{\Delta\Delta})$ \\
\tableline
 1/2    &  1/2    & (1,1),(1,2),(2,1),(2,2) & (1,0),(0,1) \\
 1/2    &  3/2    & (1,1),(1,2),(2,1),(2,2) & (0,1),(1,2) \\
 1/2    &  5/2    & (1,1),(1,2),(2,1),(2,2) & (0,3),(1,2) \\
 1/2    &  7/2    & (1,2),(2,2) & (0,3) \\
 3/2    &  1/2    & (1,1),(1,2),(2,1),(2,2) & (1,0),(2,1) \\
 3/2    &  3/2    & (1,1),(1,2),(2,1),(2,2) & (1,2),(2,1) \\
 3/2    &  5/2    & (1,1),(1,2),(2,1),(2,2) & (1,2),(2,3) \\
 3/2    &  7/2    & (1,2),(2,2) & (2,3) \\
 5/2    &  1/2    & (1,1),(1,2),(2,1),(2,2) & (2,1),(3,0) \\
 5/2    &  3/2    & (1,1),(1,2),(2,1),(2,2) & (2,1),(3,2) \\
 5/2    &  5/2    & (1,1),(1,2),(2,1),(2,2) & (2,3),(3,2) \\
 5/2    &  7/2    & (1,2),(2,2) & (2,3) \\
 7/2    &  1/2    & (2,1),(2,2) & (3,0) \\
 7/2    &  3/2    & (2,1),(2,2) & (3,2) \\
 7/2    &  5/2    & (2,1),(2,2) & (3,2) \\
 7/2    &  7/2    & (2,2) & \\
\end{tabular}
\end{table}

\begin{table}
\caption{Binding energies $B_2$ of the
$NN$ states with
total angular momentum $j$ and isospin $i$.
$B_2^L$ are the
results of the local
model and $B_2^{NL}$ are the results of the nonlocal model.}

\begin{tabular}{cccccc}
& $j$ & $i$   & $B_2^L$(MeV) & $B_2^{NL}$(MeV)    & \\
\tableline
& 1   &  0    & 3.13 & 2.14 & \\
& 0   &  1    & unbound & unbound & \\
\end{tabular}
\end{table}

\begin{table}
\caption{Binding energies $B_2$ of the
$N\Delta$ states with
total angular momentum $j$ and isospin $i$.
$B_2^L$ are the
results of the local
model and $B_2^{NL}$ are the results of the nonlocal model.}

\begin{tabular}{cccccc}
& $j$ & $i$   & $B_2^L$(MeV) & $B_2^{NL}$(MeV)    & \\
\tableline
& 1   &  1    & unbound & unbound & \\
& 1   &  2    & unbound & unbound & \\
& 2   &  1    & 0.0 & 0.141 & \\
& 2   &  2    & unbound & unbound & \\
\end{tabular}
\end{table}

\begin{table}
\caption{Binding energies $B_2$ of the
$\Delta\Delta$ states with
total angular momentum $j$ and isospin $i$. $B_2^L$ are the
results of the local
model and $B_2^{NL}$ are the results of the nonlocal model.}

\begin{tabular}{cccccc}
& $j$ & $i$   & $B_2^L$(MeV) & $B_2^{NL}$(MeV)    & \\
\tableline
& 0   &  1    & 108.4 & 159.5 & \\
& 0   &  3    & 0.4   & 0.2 & \\
& 1   &  0    & 138.5 & 190.3 &  \\
& 1   &  2    & 5.7 & unbound & \\
& 2   &  1    & 30.5 & 7.4 & \\
& 2   &  3    & unbound & unbound & \\
& 3   &  0    & 29.9 & 7.8 & \\
& 3   &  2    & unbound & unbound & \\
\end{tabular}
\end{table}

\begin{table}
\caption{Binding energies $B_3$ and separation energies $B_3 -B_2$
of the $NNN$ states with total spin $S$ and isospin $I$.
$B_2^L$ and $B_3^L$ are the
results of the local model while
$B_2^{NL}$ and $B_3^{NL}$ are the results of the nonlocal model.}

\begin{tabular}{cccccccc}
& $S$ & $I$ & $B_3^L$(MeV) & $B_3^L-B_2^L$(MeV) & $B_3^{NL}$(MeV)
& $B_3^{NL}-B_2^{NL}$(MeV) & \\
\tableline
& 1/2    &  1/2    & 5.76 & 2.63 & 6.52 & 4.38 & \\
& 1/2    &  3/2    & unbound & $-$  & unbound & $-$ & \\
& 3/2    &  1/2    & unbound & $-$  & unbound & $-$ & \\
\end{tabular}
\end{table}

\begin{table}
\caption{Binding energies $B_3$ and separation energies $B_3 -B_2$
of the $NN\Delta$ states with total spin $S$ and isospin $I$.
$B_2^L$ and $B_3^L$ are the
results of the local model while
$B_2^{NL}$ and $B_3^{NL}$ are the results of the nonlocal model.}

\begin{tabular}{cccccccc}
& $S$ & $I$   & $B_3^L$(MeV) & $B_3^L-B_2^L$(MeV) & $B_3^{NL}$(MeV)
& $B_3^{NL}-B_2^{NL}$(MeV)  & \\
\tableline
& 3/2    &  1/2    & unbound &  $-$   & 0.143 & 0.002 & \\
& 3/2    &  3/2    & unbound &  $-$   & 2.280 & 0.144 & \\
\end{tabular}
\end{table}

\begin{table}
\caption{Binding energies $B_3$ and separation energies $B_3 -B_2$
of the $N\Delta\Delta$ states with total spin $S$ and isospin $I$.
$B_2^L$ and $B_3^L$ are the
results of the local model while
$B_2^{NL}$ and $B_3^{NL}$ are the results of the nonlocal model.}

\begin{tabular}{cccccccc}
& $S$ & $I$   & $B_3^L$(MeV) & $B_3^L-B_2^L$(MeV) & $B_3^{NL}$(MeV)
& $B_3^{NL}-B_2^{NL}$(MeV)  & \\
\tableline
& 1/2    &  5/2    & unbound & $-$   & 0.630 & 0.43 & \\
& 5/2    &  1/2    & unbound & $-$   & 8.158 & 0.358 & \\
& 5/2    &  5/2    & unbound & $-$   & 0.181 & 0.04 & \\
\end{tabular}
\end{table}

\begin{table}
\caption{Binding energies $B_3$ and separation energies $B_3 -B_2$
of the $\Delta\Delta\Delta$ states with total spin $S$ and isospin $I$.
$B_2^L$ and $B_3^L$ are the
results of the local model while
$B_2^{NL}$ and $B_3^{NL}$ are the results of the nonlocal model.}

\begin{tabular}{cccccccc}
& $S$ & $I$   & $B_3^L$(MeV) & $B_3^L-B_2^L$(MeV) & $B_3^{NL}$(MeV)
& $B_3^{NL}-B_2^{NL}$(MeV)  & \\
\tableline
& 1/2    &  1/2    & 84.0 & 53.5 & 16.6 & 9.2 & \\
& 1/2    &  3/2    & 139.2 & 0.7 & unbound & $-$ & \\
& 1/2    &  7/2    & 6.3  & 0.6 & unbound & $-$ & \\
& 3/2    &  1/2    & 109.5 & 1.1 & unbound & $-$ & \\
& 5/2    &  1/2    & 39.1  & 8.6 & 9.3 & 1.9 & \\
& 7/2    &  1/2    & 31.7  & 1.2 & 7.8 & 0.4 & \\
& 7/2    &  3/2    & 35.1  & 4.6 & 9.8 & 2.0 & \\
\end{tabular}
\end{table}

\end{document}